\documentclass[rsi,amsmath,amssymb,reprint]{revtex4-1}
\usepackage[utf8]{inputenc}
\usepackage{graphicx}
\usepackage{dcolumn}
\usepackage{bm}

\usepackage[utf8]{inputenc}
\usepackage{xcolor}
\usepackage[T1]{fontenc}
\usepackage{mathptmx}
\usepackage{xspace}

\usepackage[
disable
]{endfloat}
\begin{document}

\preprint{AIP/123-QED}

\title[The Raspberry Pi Auto-aligner: Machine Learning for Automated Alignment of Laser Beams]{The Raspberry Pi Auto-aligner: Machine Learning for Automated Alignment of Laser Beams}

\author{Renju S. Mathew}
\author{Roshan O'Donnell}
\author{Danielle Pizzey}
\email{danielle.boddy@durham.ac.uk}
\author{Ifan G. Hughes}%
 \affiliation{Joint Quantum Centre (JQC) Durham-Newcastle, Department of Physics, Durham University, South Road, Durham, DH1 3LE, United Kingdom}

\date{\today}

\begin{abstract}

We present a novel solution to automated beam alignment optimization. This device is based on a Raspberry Pi computer, stepper motors, commercial optomechanics and electronic devices, and the open source machine learning algorithm M-LOOP. We provide schematic drawings for the custom hardware necessary to operate the device and discuss diagnostic techniques to determine the performance. The beam auto-aligning device has been used to improve the alignment of a laser beam into a single-mode optical fiber from manually optimized fiber alignment with an iteration time of typically 20~minutes. We present example data of one such measurement to illustrate device performance. 
\end{abstract}

\maketitle

\section{\label{sec:intro}Introduction}
Machine learning (ML) methods can discover patterns in data without requiring any assumptions about the data's structure~\cite{Murphy12}. Performing research with ML began in earnest in the 1980s~\cite{ISI:000412476200021} and by 1992 ML methods were used to, for example, create non-intuitive laser pulse-sequences for exciting rotational quantum states~\cite{Judson92}. However, it is only in the last decade or so that ML methods have begun to be used more widely in the atomic, molecular and optical (AMO) physics community.   ML techniques have  been used to create self-tuning, mode-locked lasers~\cite{Kutz:14, Grelu16,Kutz18}; for automating the production of Bose-Einstein condensation~\cite{Hush16}; and maintaining doughnut-shaped beams in scattering media~\cite{Gong16}. ML has recently been used even to create new quantum experiments: the system both learned to create a variety of entangled states and improved the efficiency of their realization~\cite{Briegel18}. Despite these advances, no work using ML for beam alignment has been found.

``Walking the beam'' is the process of aligning a laser beam using two adjustable mirrors in such a way that it will reach a specific point in space with a specific angle. It can be a laborious task to properly adjust the four knobs that control the horizontal (yaw) and vertical (pitch) angles of the two mirrors. Without making the correct sequence of adjustments, it is even possible to move further away from the goal instead of towards it. This is particularly difficult in quantum optics experiments working at low light levels. Aligning the signal into the detector is more difficult in experiments with very low light levels--such as single-photon generation from  hot thermal vapors~\cite{PhysRevA.79.033814, ISI:000286728700022, Willis:11, Ding:12, PhysRevLett.109.033601,  PhysRevA.93.043854,  Lee:16,  PhysRevA.93.053821,  doi:10.1080/09500340.2017.1377308, PhysRevA.100.033818,  Jeong:20}--that necessitate using fiber-coupled photon-detectors~\cite{Whiting_2017, PhysRevLett.122.143601, Wang:s, Mika_2020}. Because of these issues, we were strongly motivated to commission a device to automate beam alignment. Not only can this device save time and find a more optimal alignment, but alignment automation can be a helpful addition to laser-safety protocols.


In the same time frame that ML techniques have been utilized in laboratories, there has been a surge of interest in using low-cost but high-performance hardware, particularly in optics and imaging experiments~\cite{ doi:10.1063/1.4972255, doi:10.1063/1.4941068, doi:10.1063/1.5025729, doi:10.1063/1.5022973, doi:10.1063/1.4986044, doi:10.1063/1.5071447,   doi:10.1063/1.5066062}.  Here we present a device based on using stepper motors operated by a Raspberry Pi computer to control commercial kinematic mounts. By attaching a motor to the yaw and pitch knobs of each mount, the orientation of the mirror can be controlled electronically. We employ the open source machine learning algorithm M-LOOP~\cite{Hush16} to optimize the coupling of a laser beam into a single-mode optical fiber, using a commercial power meter as a detector. We call this entire set-up a ``beam auto-aligner''. 

Our design has two main advantages over the current commercial alternatives: first, our device is significantly less expensive than commercial alternatives that perform similar functions. Second, since commercial hardware and software are proprietary, the equipment must be used as a black-box, limiting room for customization. Commercial software may need to be bought separately from the hardware and need periodic updating. For our beam auto-aligner, all of the hardware drawings, computer aided design (CAD) files, and electronic schematics including a bill-of-materials are available in the Supplementary Materials. The computer software, M-LOOP, to operate the device is available on the M-LOOP website~\cite{MLOOP}.

The remainder of the paper is organized as follows: in Section~\ref{sec:hardware} we present the hardware, and discuss the calibration procedure; Section~\ref{sec:software} discusses the relevant features of the M-LOOP software; we illustrate the performance of the beam auto-aligner in Section~\ref{sec:results}; and finally conclusions are drawn and an outlook provided in Section~\ref{sec:conclusion}.

\section{\label{sec:methods}Methods}

\subsection{\label{sec:hardware}Hardware}

\begin{figure}[tbh!]
\begin{center}
\includegraphics[width=9.0cm,clip=true,trim = 0mm 0mm 0mm 0mm]{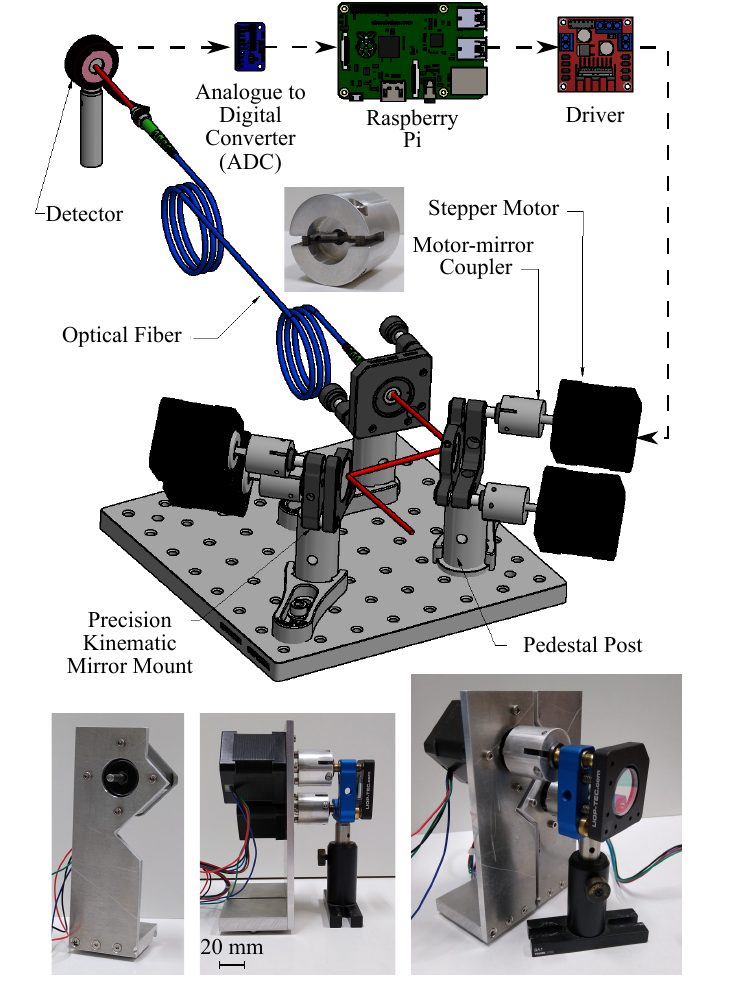}
\end{center}
\vspace{0mm}
\caption{Top: Schematic illustration of the beam auto-aligner assembly (specifically Build 2.0), showing a typical use scenario where four motors are used to control the pitch and yaw on two kinematic mirror mounts to align a laser beam into a single-mode optical fiber. The laser beam that exits fiber is directed onto a detector, with the voltage reading inputted to the Raspberry Pi via an analog-to-digital converter (ADC). Should a photon counter be used as the detector, the ADC is no longer required. The dashed black arrows convey the electrical path. For scale, the breadboard the beam auto-aligner is mounted on is 115\,mm\,x\,115\,mm. See the main text for details of the highlighted components. Connecting cables and the support mount for the motor-mirror couplers are not shown in the rendering drawing for clarity. Inset of the rendering drawing is a photograph of the motor-mirror coupler. Bottom: Photographs of the hardware. Left: L-plate motor support mount for one stepper motor. Middle and right: Illustrates how the motor attaches to the kinematic mirror mount knob via the motor-mirror coupler.
\label{fig:ML_setup}}
\end{figure}

Figure~\ref{fig:ML_setup} shows a rendering of the device. A stepper motor attaches to each kinematic mount knob via a custom-made motor-mirror coupler. The motor-mirror coupler support mount is not shown in the rendering drawing but it is shown in a photograph inset.


The computer used to run M-LOOP and to control the motors is a Raspberry Pi 4 Model B, though any version of Pi will be suitable. The Raspberry Pi has a row of GPIO (general purpose input/output) pins that allows direct communication with the motors. The device has minimal footprint, with dimensions 85~mm\,x\,56~mm\,x\,17~mm, that can be placed directly on the optical table, minimizing the footprint of the device. The output for M-LOOP to optimize is chosen to be the beam power on the output of a single-mode optical fiber. The optical beam power can be measured as a voltage on a photo-diode, a commercial power-meter, or integrated photon count. In this work, we use a homebuilt photo-diode. 

An analog-to-digital converter (ADC) is required to convert the analog output of the homebuilt photo-diode into a form usable by the Pi. The 16-bit ADS1115 ADC is used in this work. The Raspberry Pi can operate four motors, which is adequate for this investigation. However, should the user require control of more than four motors, shift registers can be used in conjunction with the Raspberry Pi to increase the number of GPIO pins of the Pi. Additionally, drivers are necessary as an intermediary between the Raspberry Pi--which output tens of milliamps--and the stepper motors--which require several amps. 


\subsection*{Calibrating the hardware}

We discuss two designs of the hardware in this section, named Build 1.0 and Build 2.0 for first and second generation hardware, respectively. Build 1.0 enabled rapid testing; however in troubleshooting the device, constraints were identified regarding the hardware design. Build 2.0 addresses the issues identified with Build 1.0; however, should the reader wish to use a modified design to suit their purpose, they should follow the troubleshooting advice provided in section~\ref{subsec:build1.0} and \ref{sec:results}.
 
\subsubsection{\label{subsec:build1.0}Build 1.0}
Initial designs used 28BYJ-48 stepper motors, driven by ULN2003 boards, in conjunction with Thorlabs standard kinematic mounts and plastic 3D printed motor-mirror couplers, allowing for rapid prototyping and testing of the M-LOOP algorithm with the motors. We investigated the reproducibility of the device by controlling the rotation of one kinematic mirror mount knob and measuring the power transmitted by the fiber.

\begin{figure}[tbh!]
\begin{center}
\includegraphics[width=8.5cm,clip=true,trim = 0mm 0mm 0mm 0mm]{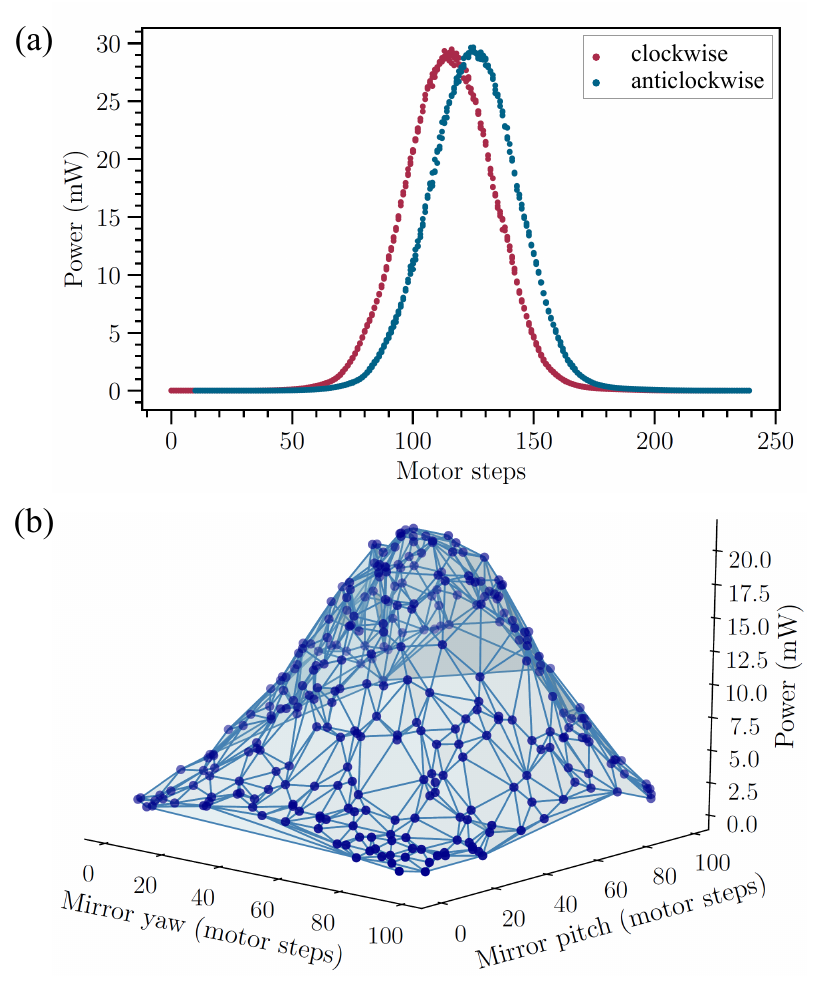}
\end{center}
\vspace{0mm}
\caption{Testing hysteresis and reproducibility of hardware designs. (a) Illustration of hysteresis in initial designs (i.e. Build 1.0). When the motor turns the kinematic mount knob clockwise from 0 to 240 motor steps (in increments of 1 step) and then turns the mirror knob anticlockwise by the same number, the mirror does not return to the same position. The amount of kinematic mount knob rotation per motor step is dependent on the motor and kinematic mount used, as discussed in the text. (b) Investigating reproducibility for two axes for Build 2.0 design. The mirror yaw and pitch orientations are initialized by hand such that the coupling efficiency into the single-mode optical fiber is maximized. The motor moves the mirror to a random orientation, as shown by the blue circles, and immediately moves back to the starting position. This is repeated many times. A landscape is mapped out. 
\label{fig:Reproduce}}
\end{figure}

Prior to initializing M-LOOP, the beam coupling into the single-mode optical fiber is optimized manually by the user using all four kinematic mirror mount knobs. Before attaching the motor via the motor-mirror coupler to the kinematic mirror mount knob of interest, the user rotates the knob anticlockwise such that the power into the optical fiber is minimal. It is at this kinematic knob position that the user defines as zero motor steps in figure~\ref{fig:Reproduce}(a). Once the device is attached, the Pi sends a command to the motor (via its driver) to turn the kinematic mirror mount knob of interest by a full 360-degree clockwise rotation in increments of one motor step. No changes are made to the other kinematic mirror mount knobs. Then the Pi sends another command to the motor to turn the original kinematic mirror mount knob in the reverse direction by the same number of steps. At each motor step the power is recorded. The results are illustrated in figure~\ref{fig:Reproduce}(a). The system exhibits hysteresis--the kinematic mirror mount knob does not return to its starting position and the coupling efficiency of the laser into the fiber changes correspondingly. We can correct for hysteresis in M-LOOP by incorporating the step-loss in the M-LOOP program. The step-loss is determined by measuring the motor step difference (i.e. the difference in $x$ value in figure~\ref{fig:Reproduce}(a)) between the curves at maximum beam power when the motor changes from clockwise to anticlockwise (and vice versa). This can be a tedious task and we found the dominating cause for hysteresis was the plastic motor-mirror coupler: these degraded and morphed over time due to the force exerted by the motor and were eventually unable to transfer
rotation to the kinematic mount knob. Subsequent motor-mirror coupler designs were machined using stainless steel and incorporated into the initial design. However, the 28BYJ-48 stepper motors did not have sufficient torque to cope with the extra load of the metal motor-mirror coupler and, as such, would frequently stall during operation.  

Although we did not pursue Build 1.0 further, one can use the methodology described to calibrate each motor-knob pair to determine the number of extra steps required for a directional change and, if necessary, to account for this when commands are sent to the motor via the Pi. Instead of accounting for step-loss, we adapted our motor-mirror coupler to eradicate hysteresis (as discussed in section~\ref{subsec:Build2.0}, i.e. Build 2.0).

\subsubsection{\label{subsec:Build2.0}Build 2.0}
Revised designs, which are included in the Supplementary Material, include a metal motor-mirror coupler and a stepper motor with greater torque and negligible backlash. We chose to use NEMA-17 stepper motors, which are a two-phase, bipolar motor, that are driven by L298N driver boards. The L298N driver has two H-bridges which are able to switch the direction of current supplied to the motors, hence can change the direction of motor rotation rapidly. To further minimize the effects of hysteresis, stable mirror mounts (SR100-100-2-BU) from Photonic Technologies (LiOp-Tec) and stable pedestals (RDS-MNI-P-75) and holding forks (RDS-MNI-HF-M) from Radiant Dye Laser are used.

To use this device, the L298N drivers require 5--35~V, but to achieve maximum motor speed a voltage of 30--35~V is advised. The Raspberry Pi requires 5~V power, and it powers the ADC and, if necessary, shift registers from this supply. The wiring for the electronic components is provided in the Supplementary Material.

The reproducibility of the revised beam auto-aligner design is investigated for two axes (i.e. two knobs on  one kinematic mirror mount). A simple test is executed whereby the mirror is moved back and forth between the starting orientation and some random orientation and the beam power coupled into the single-mode optical fiber is measured. To start, the laser beam is manually aligned into the single-mode optical fiber until maximum coupling efficiency is achieved. The motors are attached to the kinematic mount knobs and M-LOOP is engaged. The results are shown in figure~\ref{fig:Reproduce}(b). 
For the specific application of coupling light into a single-mode optical fiber, the transmitted power is expected to be a smooth, single-peaked function of mirror displacement~\cite{f2f}.  The shape of the experimentally measured surface evident in \ref{fig:Reproduce}(b) confirms that the motor-mirror set-up is working as required; the measured landscape would not have been observed in a system exhibiting hysteresis.
To summarize, Build 2.0 design addresses the hardware issues experienced in Build 1.0 as we do not measure hysteresis or mechanical drift.


\subsection{\label{sec:software}Software}



The computer software, M-LOOP, to operate the device is available on the M-LOOP website \cite{MLOOP}. The website links to their paper \cite{Hush16}, Github, and provides basic instructions on how to use the software, including how to start M-LOOP through the terminal. M-LOOP is housed on the Raspberry Pi and is controlled using terminal commands on the Pi's operating system, although it is possible to use a custom Python interface. 

M-LOOP contains several ML algorithms of which only Gaussian processes (GPs) were investigated in this work; GPs can deal effectively with uncertainty in the model because GPs are a probabilistic technique based on Bayesian inference \cite{burt2019rates}. This effectiveness with dealing with uncertainty should make
them especially useful in beams dominated by Poissonian noise, as in the case of our experiments involving single photons.

To set-up the device, the user manually couples some amount of light into the single-mode optical fiber, then attaches a motor to each kinematic knob via the motor-mirror coupler. At this stage, the user is in a position to allow M-LOOP to determine the best mirror orientations. To initialize M-LOOP, the user enters the number of runs for M-LOOP to perform. M-LOOP will try to optimize the system until it has completed $N$ runs or until it has reached the ``goal'' value which the user defines at the beginning. 

More specifically to this device application, M-LOOP is used as follows. There is some ``cost'', which is a number that the algorithm attempts to minimize--reminiscent of least-squares error analysis techniques~\cite{errors}. In this example, we are interested in maximizing the power out of the fiber, $P_\textrm{{out}}$, so we can simply set the cost to equal the negative of the power $-P_\textrm{{out}}$. M-LOOP is initialized with the boundary values of the parameter space it should explore. We have four parameters: a pitch (parameter 1, p$_{1}$) and yaw (parameter 2, p$_{2}$) for mirror 1 and a pitch (parameter 3, p$_{3}$) and yaw (parameter 4, p$_{4}$) for mirror 2. There is some unique set of values (p$_{1}$, p$_{2}$, p$_{3}$, p$_{4}$) where the coupling efficiency is at its maximum. We must initialize M-LOOP by setting the size of the parameter space, i.e. the boundary conditions (e.g. the number of motor steps M-LOOP can make both clockwise and anticlockwise from its initial position), we wish to explore. This will initially require some trial-and-error when the user is setting up the device. In section~\ref{sec:results} we guide the reader through troubleshooting the device with examples.

M-LOOP begins by outputting a file containing an initial set of parameters (p$^\textrm{{(i)}}_{1}$, p$^\textrm{{(i)}}_{2}$, p$^\textrm{{(i)}}_{3}$, p$^\textrm{{(i)}}_{4}$) for which it wants the associated cost. The Raspberry Pi sets each kinematic mirror mount knob to these values. The Pi then requests and receives the power on the detector, $P_\textrm{{out}}$, in units of voltage. The negative of this number ($-P_\textrm{{out}}$) is written to a file containing the value of the cost. M-LOOP waits until it detects this file and writes a new file containing new parameters and the process repeats. 

\begin{figure}[tbh!]
\begin{center}
\includegraphics[width=7.5cm,clip=true,trim = 0mm 0mm 0mm 0mm]{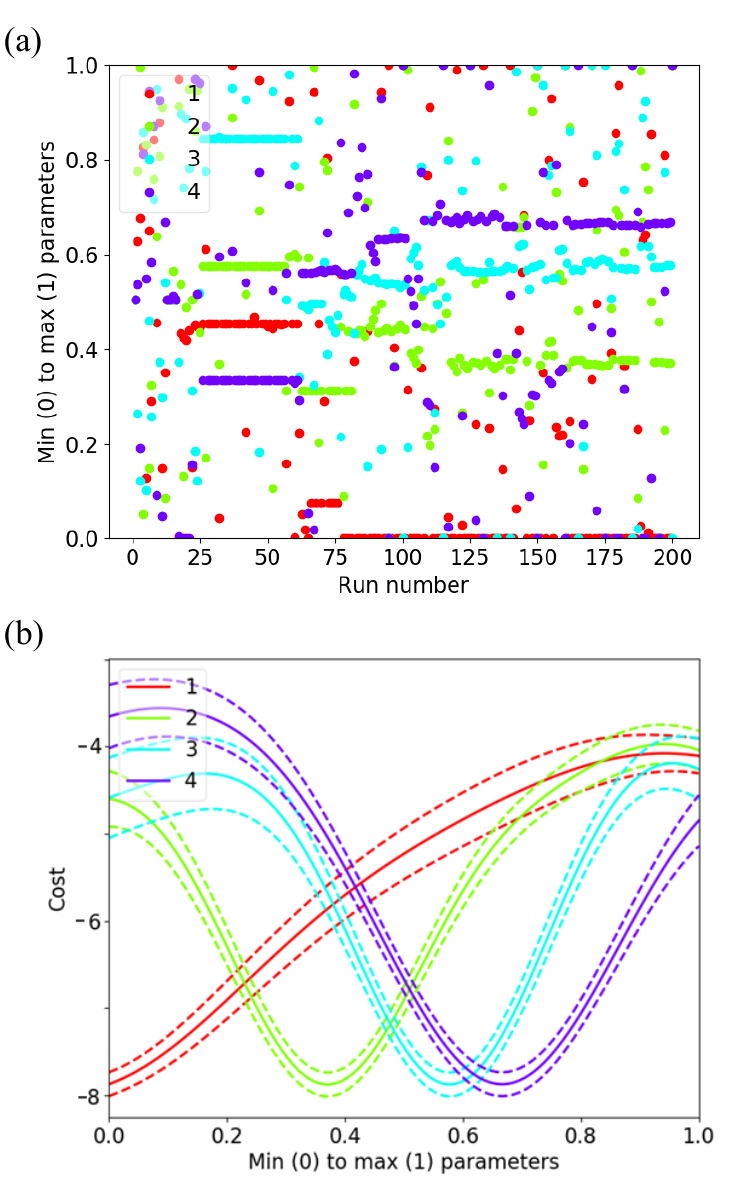}
\end{center}
\vspace{0mm}
\caption{Direct output from the M-LOOP algorithm. The parameter values labelled 1, 2, 3, and 4 in the legend are red, green, blue and purple in the figures, respectively. The parameters correspond to the pitch (``1'') and yaw (``2'') for mirror 1 and the pitch (``3'') and yaw (``4'') for mirror 2. (a) The $y$-axis is the normalized number of turns on a kinematic mirror mount knob where 0.5 means no change from the original position. After a certain number of runs, M-LOOP homes in on the best parameters. For this dataset, the best value for parameter 1 (red) lay outside the range the user allowed M-LOOP to explore. (b) With a sufficient number of runs, M-LOOP builds an internal model of the parameter landscape. The dashed lines indicate error boundaries. Smooth curves indicate that the device is functioning as expected. For this dataset, the minimum of parameter 1 lay outside the range the user allowed M-LOOP to explore. 
\label{fig:Direct_MLOOP_output}}
\end{figure}

The initial set of parameter values is random. However, on each run of the above, M-LOOP begins to build an internal model of the parameter landscape. From this, it begins to test the parameters' values most likely to minimize the cost. M-LOOP finishes when one of the three halting conditions is met: (1) The maximum number of runs has been reached. (2) The minimum cost value has been reached. (3) A certain number of runs has elapsed without a lower cost being found. The user sets these values before the first run.  
It is always a trial-and-error approach to  set the initial boundary conditions when initializing M-LOOP. Typically, if the user sets the boundary conditions to allow a full 360-degree rotation of the kinematic mount knob, M-LOOP is able to find optimum coupling efficiency. The relationship between the number of motor steps and the amount of kinematic mirror knob rotation depends on the particular kinematic mount and motor used. For Build 1.0, which uses standard Thorlabs kinematic mounts, we measure 360 steps per 360-degree rotation. For Build 2.0, which uses LiOp-Tec mirrors, we measure 200 steps per 360-degree rotation. In practice, it is always the number of steps that are the relevant quantity in the M-LOOP algorithm. Troubleshooting the best M-LOOP boundary space is not something we can fix for the user, like we can for the hardware design. We can, however, give guidance to the user on how to address the software issues by using M-LOOPs direct outputs, an example of which is shown in figure~\ref{fig:Direct_MLOOP_output}. Note that we provide further troubleshooting advice in section~\ref{sec:results}.

Figure~\ref{fig:Direct_MLOOP_output} is a direct output from the M-LOOP algorithm. The user does not observe the plotting of this figure in real-time; it is only once M-LOOP has met one of the three halting conditions that the user obtain these results. Careful
study of the output gives insight into what M-LOOP is doing at each run and gives indications on how to troubleshoot. For example, we can see that for figure~\ref{fig:Direct_MLOOP_output}(a), for the first 25 runs a random set of parameters is tried. This is typical, though the number of runs during this random search varies. By run 25, M-LOOP has homed in on an optimum set of parameter values as shown by the plateaus for all four parameters. It continues testing the nearby parameter space but regularly tries completely different parameters to ensure that it is not stuck in a local cost minimum~\cite{errors}. Indeed, by run 70, it has discovered a different set of optimal parameters. However, by run 77, parameter 1 (red) is at the very edge of its constraint. The algorithm ``wants'' to explore values below 0 but our initial boundary conditions do not allow it. This suggests that we have set the boundary of the parameter space for parameter 1 too narrowly and that we should re-run the whole process with this boundary
expanded. This is also evident in figure~\ref{fig:Direct_MLOOP_output}(b), which shows the predicted landscape of cost against parameter value for the
same experimental set shown in figure~\ref{fig:Direct_MLOOP_output}(a). The minimum of the parameter 1 curve (red) seems to lie just outside of the range that the user has allowed M-LOOP to investigate as can be seen if we imagine extrapolating the red curve into the
region of the negative $x$-axis. Nevertheless, the smooth curves for each of the four
parameters demonstrate that our program is working as expected. M-LOOP takes
approximately 20 minutes from start to finish for 200 runs.


\section{\label{sec:results}Results}

\begin{figure*}[p]
\begin{center}
\includegraphics[width=8.5cm,clip=true,trim = 0mm 0mm 0mm 0mm]{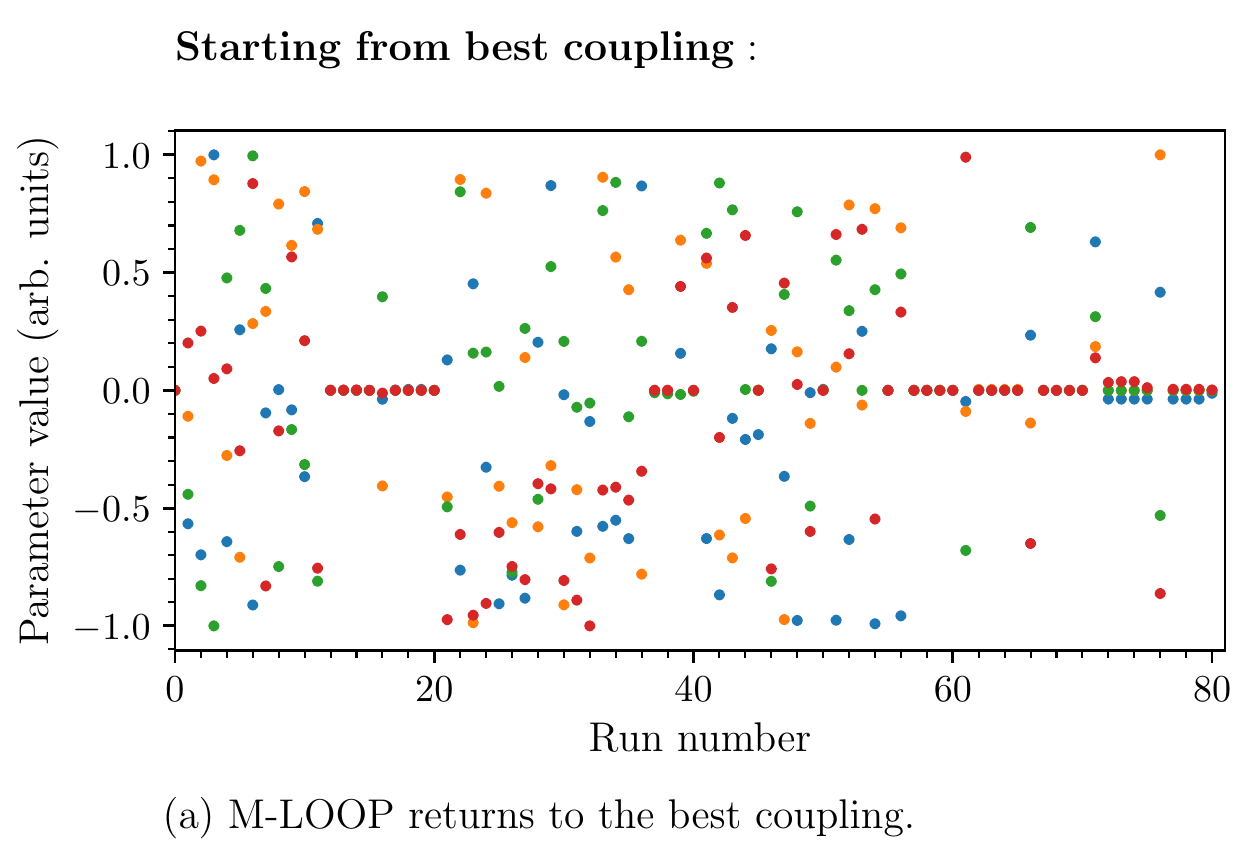}\hspace{0.5cm}
\includegraphics[width=8.5cm,clip=true,trim = 0mm 0mm 0mm 0mm]{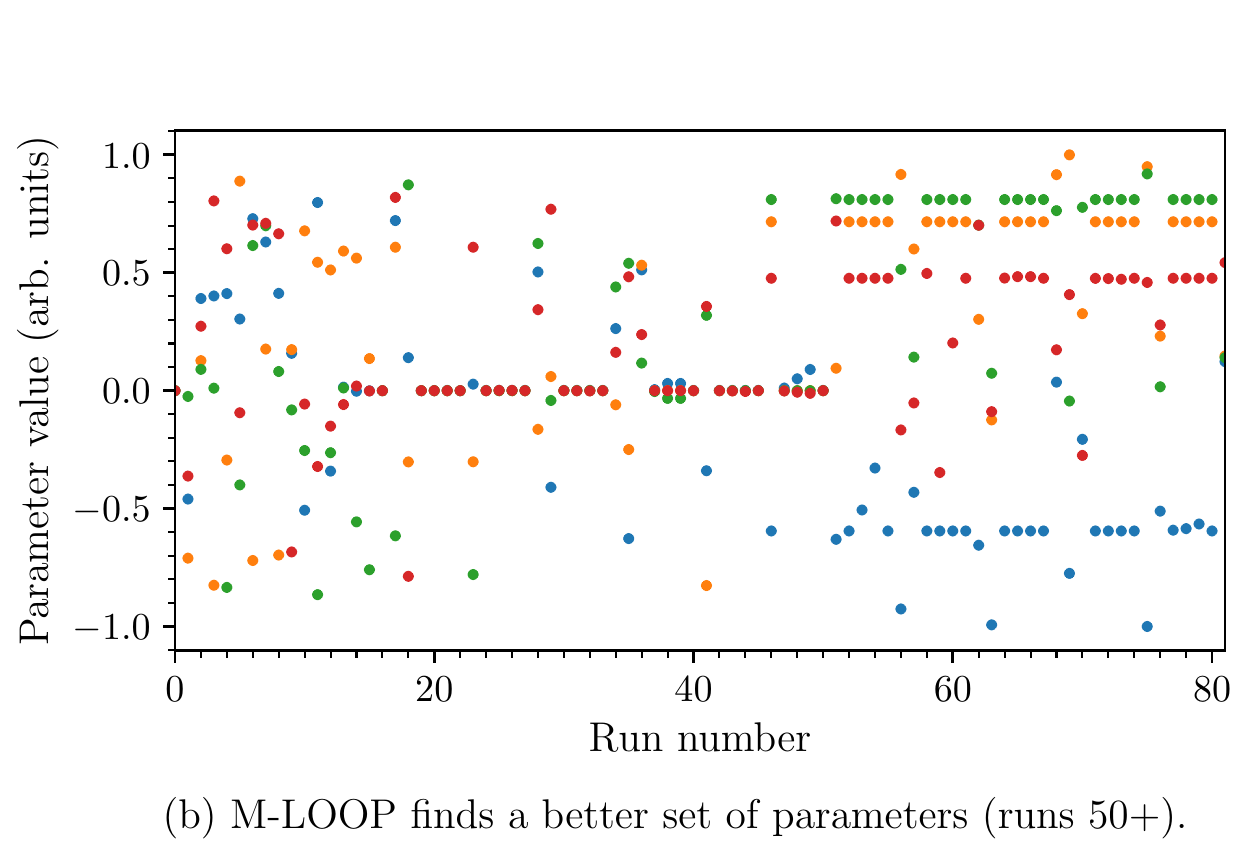}\vspace{0.25cm}
\includegraphics[width=8.5cm,clip=true,trim = 0mm 0mm 0mm 0mm]{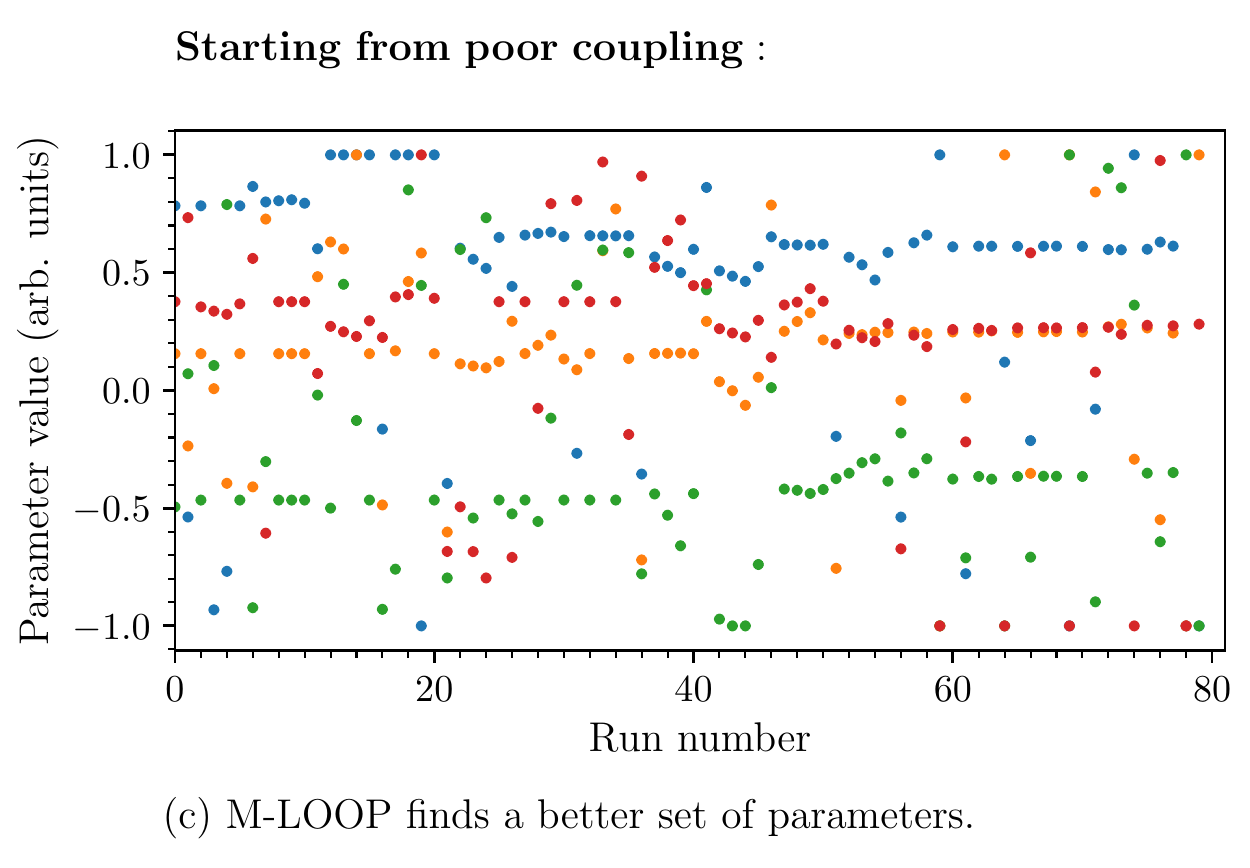}\hspace{0.5cm}
\includegraphics[width=8.5cm,clip=true,trim = 0mm 0mm 0mm 0mm]{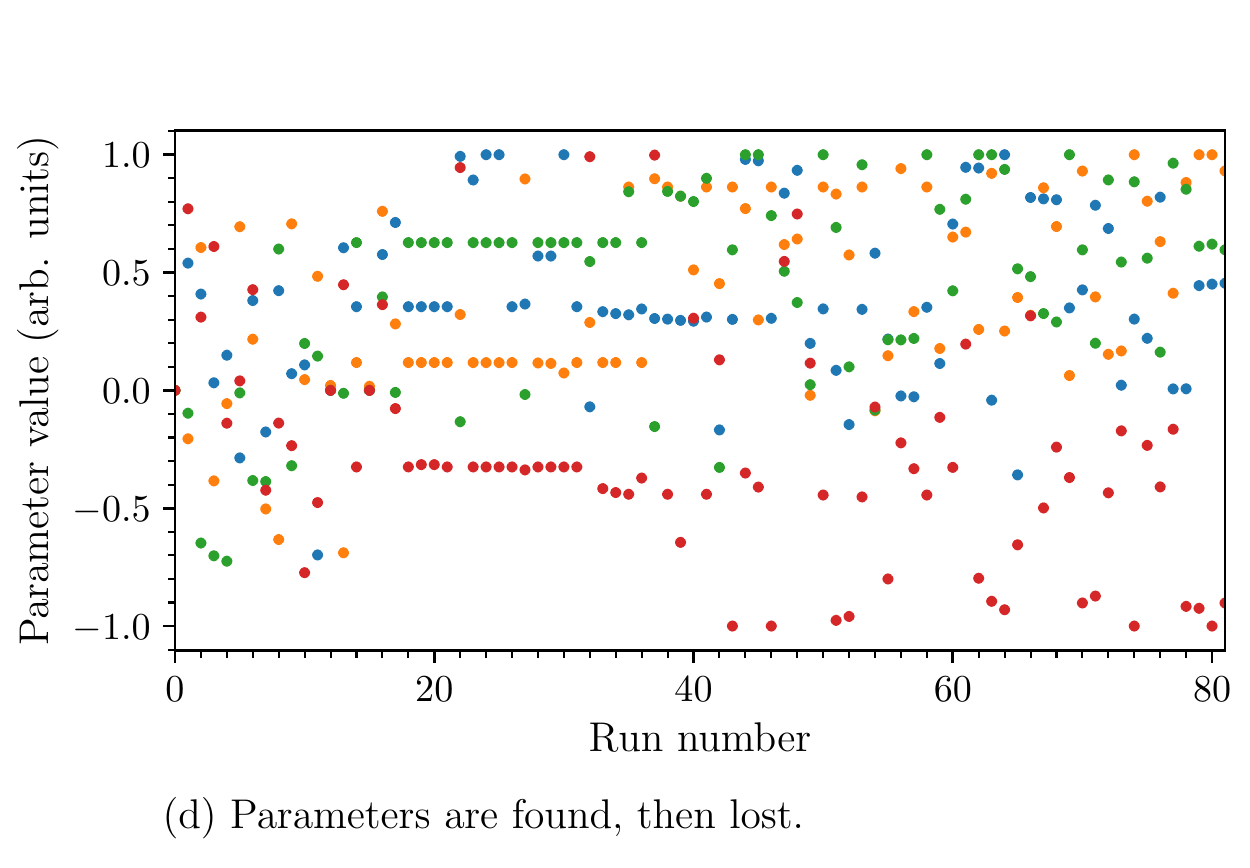}
\includegraphics[width=8.5cm,clip=true,trim = 0mm 0mm 0mm 0mm]{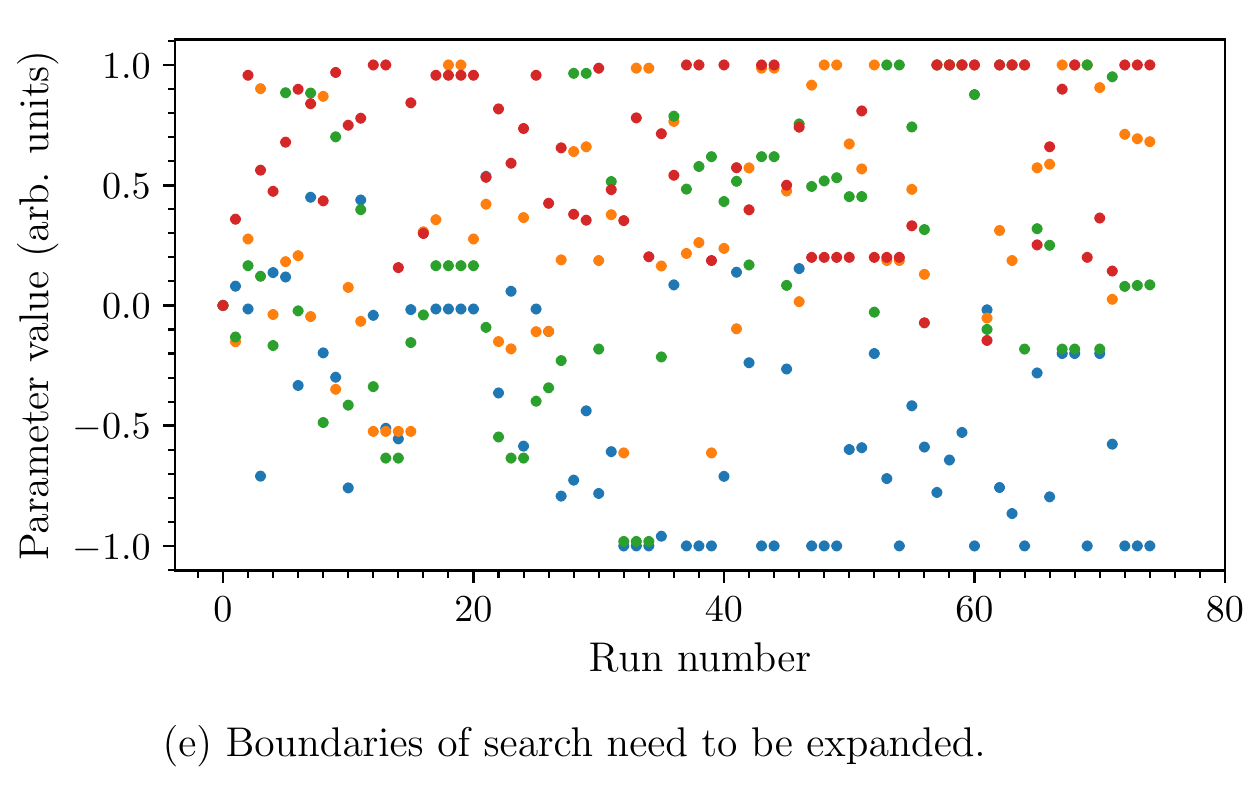}\hspace{0.5cm}
\includegraphics[width=8.5cm,clip=true,trim = 0mm 0mm 0mm 0mm]{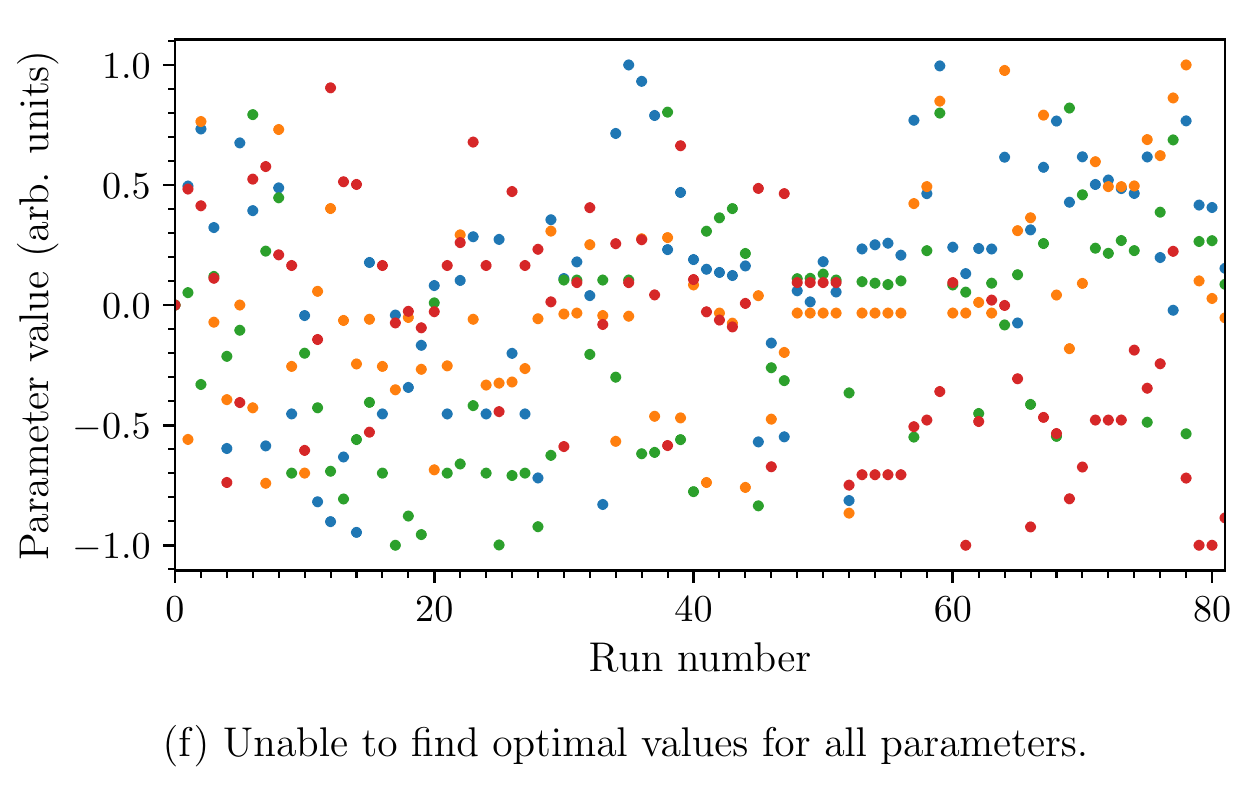}

\end{center}
\vspace{0mm}
\caption{Starting from best (manually aligned) fiber coupling (top row),  (a)-(b) illustrates M-LOOP in action. Starting from poor fiber coupling (middle and bottom rows), (c) illustrates that M-LOOP can be effective even when starting from poor fiber coupling and (d)-(f) help the user troubleshoot M-LOOP when there are issues with mechanical hardware or setting up the device. Build 1.0 hardware is used throughout all cases to illustrate issues that might arise. In all examples, parameters p$_{1}$, p$_{2}$, p$_{3}$, p$_{4}$ are represented by blue, red, orange, and green, respectively.
\label{fig:ML in operation}}
\end{figure*}

In the remainder of the paper, we illustrate the use of the beam auto-aligner and how to troubleshoot M-LOOP. A diode laser with 780~nm wavelength is manually coupled into a single-mode optical fiber such that the detector measures a non-zero amount of light (not necessarily the optimum coupling). The motors are attached using Build 1.0 design to make hardware issues more evident in the examples (this is particularly useful should the user wish to commission their own hardware design). Once set-up and the electrical power is engaged the user initializes M-LOOP. For ease of direct comparison between initial starting conditions, the parameter values are normalized in figure~\ref{fig:ML in operation} so that the parameters only ever take a value between -1 and 1.

Figure~\ref{fig:ML in operation}(a) and \ref{fig:ML in operation}(b) show what happens when we start the machine learning after manually aligning the fiber to what we, the user, believe to be the best coupling efficiency. In figure~\ref{fig:ML in operation}(a), as expected, M-LOOP does a random exploration of the parameter space for 10 runs and quickly finds that the original location (0, 0, 0, 0) is optimal. M-LOOP nevertheless tests to see if there are any better values for these parameters until on run 55 it decides that indeed (0, 0, 0, 0) is the best case. Although M-LOOP continues to test further, it does not find better parameters. The fact that the points from runs 70 to 80 are not exactly overlapping is an indication of the mechanical drift between runs (i.e. there is a problem with the coupling between the motor and the kinematic mount knob, which is typical only of our Build 1.0 design). Figure~\ref{fig:ML in operation}(b) is the case where the auto-aligner works exactly as it was designed to do. As before, we start at a position we think is the best. Between runs 15 to 50 M-LOOP agrees with the user until it discovers an even better set of parameter values.

In the remaining figures (figures~\ref{fig:ML in operation}(c)-(f)), we do not start at a position we think gives the best coupling efficiency. Instead, we make sure there is enough light entering the fiber such that the photo-diode detects some light. Figure~\ref{fig:ML in operation}(c) shows M-LOOP discovering a good set of parameters after 60 runs. The optimization is not as straightforward for M-LOOP as it is for the case shown in figure~\ref{fig:ML in operation}(b), evident by the large parameter range explored for all four kinematic mirror mount knobs. But what this does demonstrate is that M-LOOP can optimize from what the user believes is sub-optimal coupling, so if there is any mechanical drift of the device hardware or if thermal laboratory fluctuations induce a change in fiber coupling, M-LOOP should be able to account for such instabilities. This is exactly what the beam auto-aligner is intended to do. 

Figure~\ref{fig:ML in operation}(d)-(f) show failure cases where M-LOOP has to be reset and run again. These cases are explicitly for Build 1.0 to illustrate what one might expect should the device hardware need calibrating. In figure~\ref{fig:ML in operation}(d) M-LOOP seems to find a good set between runs 15 to 20 but is not sure of the correct value for the parameter corresponding to blue. We find in these cases that no changes need to be made to any M-LOOP configuration and simply re-running the procedure will result in a good set being found. In figure~\ref{fig:ML in operation}(e), however, it is evident that M-LOOP wants to look outside the boundary conditions we have given it for the parameters corresponding to both red and blue (since these are both at the extremes of the parameter value). In this case, we must re-run M-LOOP having expanded the boundaries within which it is allowed to search. It is not so clear in figure~\ref{fig:ML in operation}(f) why M-LOOP has been unable to find the optimal parameters. The solution is usually to restart the whole procedure but set the starting parameters to some sensible new values, e.g., the parameter set at 55 for figure~\ref{fig:ML in operation}(f).

Using Build 2.0 hardware, M-LOOP was able to get the same, if not slightly better, coupling efficiency than an experienced experimenter might achieve starting from poor fiber coupling. However, the worse the starting conditions (i.e. the poorer the coupling efficiency), the more input that is required from the user to set-up the device to begin with. The purpose of this device is more for continual auto-optimization, rather than using the device to couple into a fiber from scratch. The reader should note that once the M-LOOP boundary space has been set during the initial device calibration, no further software troubleshooting, or input, is required from the user. To incorporate this device into a working laboratory environment, the user would manually couple a non-zero amount of light into an optical fiber, attach the motors and run M-LOOP for its initial optimization cycle. Once M-LOOP has finished and has found the mirror knob parameter values for optimum coupling, the user can turn off the power to the motors, Raspberry Pi and associated electronics; the mirror position and angle will remain as it was at the end of the M-LOOP optimization cycle. The user can then use the device for continual auto-optimization, e.g. the device could be set to perform auto-alignment at a set time every morning before an experiment is carried out. This continual auto-optimization cycle should operate more quickly than the initial optimization cycle, as it is much easier to return a signal to maximum once the signal has been found because only a small, local parameter space needs to be explored. Furthermore, remote alignment could also be easily performed as the Pi can be controlled via the intranet using the commonly used Secure Shell (SSH) protocol, which is inbuilt into the Pi's Linux operating system.

The device is not limited to optimizing the laser power into an optical fiber, other applications include overlapping beams in a pump-probe experiment, where the peak-to-peak voltage of the signal would replace the photo-diode voltage of the beam power measured on the output of the optical fiber. There is the issue of scaling, however. The more parameters M-LOOP has to optimize, the longer it will take. Typically for our beam auto-aligner device, it took 20 minutes for M-LOOP to optimize four parameters (i.e. two mirrors having two control knobs each). If we were to use this device in a 4-wave mixing experiment, whereby the signal depends on the overlap of three laser beams, M-LOOP would need to optimize twelve parameters (i.e. each laser  beam having two mirrors for alignment control, each mirror having two control knobs). For Gaussian processes, the machine algorithm investigated in this work, the computational times scale cubically with the number of data \cite{burt2019rates}; this is not to say the device will not work to include more than four parameters, only that we have not investigated other machine learning algorithms to facilitate this. 




\section{\label{sec:conclusion}Conclusion}

In conclusion, we have demonstrated how the laborious and time-consuming task of manually aligning beams into single-mode optical fibers can be automated using our beam auto-aligner, which implements the machine learning algorithm M-LOOP. The device is not limited to optimizing laser beam power into optical fibers; other applications include overlapping beams in a pump-probe experiment or aligning high-power dipole trap laser beams with an atomic cloud in cold-atom experiments, to name a few. The intended use of the device is for continual auto-optimization, rather than using the device to align laser beams from the initial set-up. Additionally, due to the functionality of the Pi, remote alignment can easily be performed. The device has some advantages over current commercial systems; the components~\footnote{Throughout this article we have provided details of the commercial components
that we have used, in order to assist the reader who wishes to duplicate our system. Components from other manufacturers may  deliver
similar, or better, performance} are comparatively cheap and the operating software M-LOOP is open source; however the device can suffer from hysteresis if the mechanical hardware is not appropriately designed or fabricated. In spite of this, we have provided the reader with advice on how to troubleshoot their own device.  

\section{\label{sec:SuppMat}Supplementary Material}

In addition to M-LOOP's GitHub repository \cite{MLOOP}, see supplementary material for hardware CAD drawings, parts list (including a bill-of-materials), and electronic schematics.
\\

\section{\label{sec:ack}Acknowledgments}

We gratefully acknowledge Nicholas Spong for providing support getting the project off of the ground, Clare Higgins for testing the device and for editorial discussions, Thomas Cutler and Lina Marieth Hoyos-Campo for editorial discussions, Aidan Hindmarch for sourcing equipment, and EPSRC (Grant No. EP/R002061/1) for funding.

\section{\label{sec:app}Data Availability}

The data that support the findings of this study are openly available in DRO at https://doi.org/[doi], reference number [reference number]

\bibliography{REFERENCES}
\end{document}